\documentclass[namedreferences]{solarphysics}

\usepackage[hyperref,optionalrh,showbiblabels]{spr-sola-addons} 
\usepackage{graphicx}        
\usepackage{color}           
\usepackage{breakurl}        

\def\P{\mathsf{P}}
\def\e{\mathsf{E}}

\newcommand{\aap}{    {\it Astron. Astrophys.}}

\newcommand{\apj}{    {\it Astrophys. J.}}

\newcommand{\jgr}{    {\it J. Geophys. Res.}}
\newcommand{\mnras}{  {\it Mon. Not. Roy. Astron. Soc.}}

\newcommand{\solphys}{{\it Solar Phys.}}

\chardef\us=`\_

\begin{document}

\begin{article}
\begin{opening}

\title{Estimating the Maximum Intensities of Soft X-ray Flares Using Extreme Value Theory}

\author{V.~\surname{De la Luz}$^{1}$\sep
        E.P.~\surname{Balanzario}$^{2}$\sep
        T.~\surname{Tsiftsi}$^{2}$
}
\institute{$^{1}$ Conacyt - SCiESMEX - LANCE, Instituto de Geof\'isica, Unidad Michoac\'an, Universidad Nacional Aut\'onoma de M\'exico, Morelia, Michoac\'an, 58190, M\'exico.
  email: \url{vdelaluz@igeofisica.unam.mx}\\  
  $^{2}$ Centro de Ciencias Matem\'aticas, Universidad Nacional Aut\'onoma de M\'exico, Apartado Postal 61-3 (Xangari), Morelia  Michoac\'an, M\'exico.\\
	}

\runningauthor{De la Luz et al.}
\runningtitle{Maximum Intensities Flares}

\begin{abstract}
Solar flares are one of the most energetic events in the solar system, their impact on Earth at ground level and its atmosphere remains under study. The repercussions of this phenomenon in our technological
infrastructure includes radio blackouts and errors in geopositional and navigation systems that are considered natural hazards in ever more countries. Occurrence frequency and intensity of the most energetic solar flares are been taken into account in national programs for civil protection in order to reduce the risk and increase the resilience from Space Weather events. In this work we use the statistical theory of extreme values as well as other statistical methods in order to asses the magnitudes of the most extreme solar flare events expected to occur in a given period of time. We found that the data set under study presents a dual tail behaviour. Our results show that on average we can expect one solar flare
greater than X23 each 25 years, that is to say, one such event each two solar
cycles.
\end{abstract}
\end{opening}

\section{Introduction}
     \label{S-Introduction} 


One of the most energetic stellar 
activities is produced by 
flares \citep{Katsova2018}. In the case of the Sun the flares occur over the solar surface, mainly in active regions \citep{2017ApJ...834...56T}. In these areas, the magnetic structures can produce very large amounts of energy that
     are released in the magnetic reconnection process \citep{2010ARA&A..48..241B}.
This magnetic reconnection has the ability to perform a transformation of energy between magnetic and kinetic energy. The full process that involves solar reconnection and its changes in energy distribution of the system's surroundings is known as a solar flare \citep{1960MNRAS.120...89G}.

When the charged particles are 
accelerated by a flare, they can 
produce an amount of electromagnetic 
waves at all wavelengths of the spectrum 
(in the most energetic, even gamma rays 
\citep{2014ApJ...787...15A}). Their 
intensity and dynamic evolution depends 
on their interactions with their environments \citep{1993ApJ...411..362C}. 
The emission at soft X-ray wavelengths produced during a solar flare has been used as a measure of the intensity of the whole complex solar flare process \citep{2017SoPh..292..144C}.
The X-ray flare classification shows
indirectly the amount of energy released
during a solar flare event \citep{goes}. The statistical
study of occurrences of solar flares
using X-ray classification allows us
to estimate the magnitudes of the most
extreme intensities a solar flare
can reach in a given period of time
\citep{Koons,Riley}. 
Knowing the magnitude
of the most extreme events is of great
interest in the context of Space Weather
studies due to the fact that the 
protocols of civil protection related 
with this natural phenomenon
need information regarding
the worst expected scenarios \citep{SWE:SWE20303}.

The WSPC/NOAA has regular records of solar
flares since 1975 until the present day.
From 1975 up to 1978 the records used H$\alpha$
observations, but after 1978 a set of
GOES X-ray detectors were used in order
to record the solar flare events. SWPC
has made efforts in order to maintain the
X-ray solar flare classification constant
in between changes of detectors installed
in the space crafts and they released an
homogeneous public list of solar flare
records from 1 September 1975 up to
28 June 2017\footnote{The data can be found in \url{https://www.ngdc.noaa.gov/stp/space-weather/solar-data/solar-features/solar-flares/x-rays/goes/xrs/}}.

In this work we use the methods of the
statistical theory of extreme values  
(see \cite{Castillo,Coles,deHaan})
as well as \emph{ad hock}
statistical methods in order to asses 
the magnitudes of the most extreme solar
flare events that can be expected to occur in
the following years. For this purpose we use
the NOAA X-ray classification data set
from 5 November 1975 to 9 October 2017.

In Section \ref{statistic} we set up the statistical problem. Section \ref{rawdata} describes the raw data. In Section \ref{tail} we study the tail behavior of the distribution of the solar flare intensities. We do this by fitting five different models for the distribution under study.

\section{Setting up the statistical problem}\label{statistic}
We are interested in estimating the
probability of a solar flare intensity
exceeding a threshold $x$. For this
end, we need information about the 
upper tail of the distribution of values
$X$ of a solar flare intensity. Denote by
$X_{(n)}$ the maximum of a
sample $X_1,\cdots,X_n$ of size $n$
of the random variable $X$ and write
\[
p_n(x)\>=\>
\P\big\{X_{(n)}>x\big\}.
\]
By the tail of the distribution of
$X_{(n)}$ we mean the behavior of
$p_n(x)$ as $x\to\infty$.
For a fixed $n$, the
slower $p_n(x)$ tends to
zero as $x\to\infty$, the greater
the probability of observing the
event $\big\{X_{(n)}>x$\big\}. An inspection
of the raw data to be depicted
shortly leads us to expect a tail
behavior of the form
\begin{eqnarray}\label{ec1}
p_n(x)\>\propto\>
x^{-\alpha}
\end{eqnarray}
for a suitable positive $\alpha$ to be
estimated from the data set. Since
$X$ and $X_{(n)}$ belong to the
same domain of attraction of a
max-stable distribution, then
the value of $\alpha$ does not
depend on $n\in\mathbb{N}$ (see
\citep{Leadbetter}, page 15). Notice also
that the smaller the value of $\alpha$,
the slower $p_n(x)$ tends to
zero as $x\to\infty$. Therefore, the
smaller the value of $\alpha$, the
larger the probability that $X_{(n)}$
assumes a large value.

By using the collected
data set corresponding for the
intensities $X$ of the observed solar flares, we
approach the task of estimating the
parameter $\alpha$ which determines the
tail behavior of the distribution
of $X_{(n)}$ by each of the following
methods.
\begin{enumerate}
\item By fitting a log-logistic 
model to small random subsets of the whole data set.
\item By the block approach of the 
extreme value theory, where we fit a
Fr\'echet distribution to the 
maxima $X_{(n)}$ of blocks of size $n$ from the
data set (\cite{Coles}, chapter 3).
\item By the threshold approach of the 
extreme value theory, where we fit a
Pareto distribution to the excesses 
$X_{(n)}-u$ for a given $n$ and
a threshold value $u$ (\cite{Coles}, chapter 4).
\item By fitting to the entire data
set a probability distribution composed
of a log-logistic part and a Pareto part.
\item By fitting to the entire data
set a probability distribution composed
of a log-logistic part and a tempered
Pareto part.
\end{enumerate}
It will turn out that
each of these methods of estimating
$\alpha$ have advantages and disadvantages.
By comparing the estimates for $\alpha$ obtained
by each of the above methods, we are in
a position to offer a wide perspective from
which one can gain an informed understanding of the
likelihood of observing an extremely large
solar flare.

\section{Description of the raw data}\label{rawdata}

\begin{figure}[b]
\begin{center}
\includegraphics[width=1.0\textwidth]{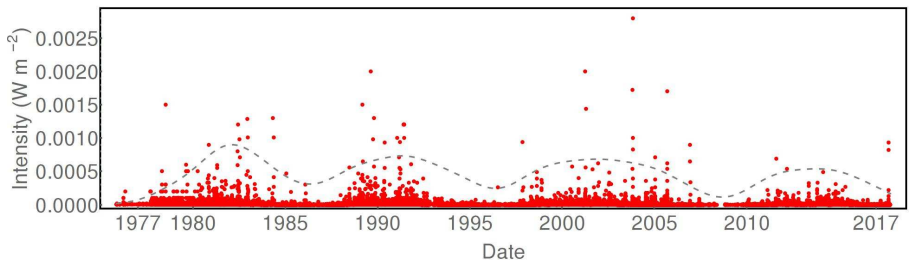}
\caption{The raw data. The
\textit{dashed gray line} represent the
density of event occurrence.}
\label{fig1}
\end{center}
\end{figure}

In Figure~\ref{fig1} we can see a
graphical representation the magnitude
(in W m$^{-2}$)
of the observed solar flare
intensities at a given time from
5 November 1975 to 9 October 2017. 
The data set here analyzed consists
of $77\,370$ data points. In columns 
with labels ``Range" and ``Data" of 
Table~\ref{t1} we further describe the
distribution of the solar flare
intensities. We
notice that an overwhelming majority of the
intensities are not greater than
0.0001 (W m$^{-2}$), this peak flux value represents an X1 GOES-class flare. Only a relatively few observations
(0.64\%) lie
above the value 0.0001 (W m$^{-2}$). 
On the other
hand, it is apparent that these extreme
observations are much greater than
0.0001 (W m$^{-2}$). 
Having few observations
much greater than the body of
the data points suggests a
behavior for the tail distribution
as depicted in Equation~\ref{ec1}
with a positive $\alpha$.
The maximum
observed value equals 0.0028 (W m$^{-2}$). 
See Table~\ref{t2}, where we report
a list of the intensities of 
the most severe events
in our data set together with 
their dates of occurrence.

The gray line in Figure~\ref{fig1}
is proportional to the density
$d(t)$ of the number of events
per unit time. Notice in particular
that this density is not uniform
over time.
The maximum value for the
density $d(t)$ occurs at
$t=1982.08$ and this maximum
equals 0.045, which represents
9.59 events per day. The minimum
of $d(t)$ occurs at $t=2008.7$ and
this minimum equals 0.0058, which
corresponds to 1.23 events per day.
The average number of events over
the whole period of time here considered
equals 5.05248 flares per day.

We computed the autocorrelation
function for the raw data for 
distinct lags.
Small values for the autocorrelations for 
positive lags were found, in spite 
of the fact that
large values for solar flare intensities
tend to occur in clusters (see for example,
entries 12 to 15 in Table~\ref{t2}). 
The small values for the autocorrelations
suggest, at least in a
first approximation, that the data set comes
form independent observations
of the random variable $X$.
Having a data set consisting
of independent observations of
a random variable allows us to
apply with confidence distinct
results from the theory of probability
and statistics, for example, the
procedures for the estimation
of the parameters of a proposed
model.

\section{Tail behavior}\label{tail}

In the second column of Table~\ref{t3}
we report the estimates for the 
tail index $\alpha$, defined by
Equation~\ref{ec1}, as given by 
the models considered
in this work. In the third column of
the table, we report 95\% confidence intervals
for $\alpha$. In each case,
the estimation for 
$\alpha$ and the confidence interval
were computed from a sample of size
given in the fourth column of Table~\ref{t3}.
In the next subsections
we comment some details about the
estimation procedures reported
in Table~\ref{t3}.

\subsection{Log-logistic model}

Among the five models considered,
the log-logistic is the most simple
minded, but even if it 
only provides us a rather naive way
to estimate the probabilities of
large solar flare intensities,
the log-logistic model provides us
a rough model for the phenomenon we
are interested in.
Now, if we model a solar flare intensity
$X$ by a log-logistic distribution, then
we have
\begin{eqnarray}\label{e2}
\P\big\{X>x\big\}\>=\>
\frac{1}{1+(x/\sigma)^\alpha}
\>\sim\>\Big(\frac{x}{\sigma}
\Big)^{-\alpha}
\end{eqnarray}
when $x$ is large, and
where the first equality in 
Equation~\ref{e2} defines the
log-logistic distribution.
When fitting the log-logistic
distribution to our data set,
the parameters $\alpha$ and $\sigma$
in Equation~\ref{e2}
were estimated by the maximum
likelihood method.
However, for the
log-logistic model we performed the
estimation of parameters, not from
the full data set consisting in
$77\,370$ points, but from a smaller
random sub-sample of size $500$,
as reported in Table~\ref{t2}.
For the null hypothesis stating
that the chosen sub-sample of
size 500 comes from a log-logistic
distribution, the Pearson chi
squared statistic for the
evaluation of the goodness of fit
has a $p$-value of $0.066$.
Thus, the null hypothesis is
not rejected, but one should
remark that the $p$-value decreases
quickly to zero when the sub-sample
size increases.
By applying both, the method
of normal approximation and the
bootstrap method\footnote{
See \cite{DiCiccio} and
\cite{Diaconis} for a
description of the bootstrap method.
}, we found that
$(1.25, 1.45)$ is a 95\% confidence
interval for $\alpha$.
When we substitute the estimated
values $\hat{\alpha}=1.3492$ and
$\hat{\sigma}=1.6240\times10^{-6}$,
we obtain from Equation~\ref{e2} that
\begin{eqnarray}\label{e3}
\P\big\{X>x\big\}\>\sim\>
\frac{1.5439\times10^{-8}}{
x^{1.3492}}
\end{eqnarray}
for large values of $x$.
We will come back to the log-logistic
model when we analyze our data
set by plotting it on log-log scale.

\subsection{The block approach}

Here we investigate the distribution
of $X_{(n)}=\max\big\{X_1,\cdots,X_n\big\}$
for a given value of $n$, where $n$ is
the size of a block in the block
approach. In view of
the estimated value for $\alpha$ obtained
for the log-logistic model, then according to
the Tippet-Fisher theorem from the
extreme value theory, we may expect that
$X_{(n)}$ has a Fr\'echet distribution
(\cite{deHaan}, page 6).
Equivalently, we expect that $X_{(n)}$
has a distribution given by
\begin{eqnarray}\label{e4}
{\rm G}_{\rm ev}(x)
=\exp\Bigg\{-\Bigg[1+\xi
\Big(\frac{x-\mu}{\sigma}\Big)
\Bigg]^{-1/\xi}\Bigg\}
\end{eqnarray}
with a positive value for the
shape parameter $\xi$.

Assuming that the events are
distributed uniformly over
time (with a rate of 5.05248
events per day), then 
a block size of $n=1\,300$
corresponds to approximately
257 days of solar activity.
This gives us a sample
of maximum values of size 59 
from which we
can estimate the parameter
$\alpha$, which in the model
given by Equation~\ref{e4}
is given by $\alpha=1/\xi$.

With the above value for $n$,
the parameters $\mu$, $\sigma$
and $\xi$ where estimated by the
maximum likelihood method. Using
this method we found that
$\hat{\mu}=3.313\times10^{-4}$,
$\hat{\sigma}=2.854\times10^{-4}$
and $\hat{\xi}=0.4145$. From this
estimate for $\xi$ we obtain
that $\hat{\alpha}=2.4122$.
By using the method of normal 
approximation, we obtained
a 95\% confidence interval
for $\xi$ given by
$(0.179, 0.650)$. From this
we can obtain a confidence
interval for $\alpha$, as
reported in Table~\ref{t3}.
For the null hypothesis
stating that the 59 maximum
values of blocks of size 1\,300
come from a Fr\'echet distribution
(equivalently, from the
distribution given by Equation~\ref{e4})
the Pearson chi squared statistic has
a $p$-value of 0.918. Thus, we have
a good fit of the maxima to the 
estimated model.

\subsection{The threshold approach}

In order to apply the threshold approach
to the estimation of $\alpha$, it is
necessary to form a set of excesses
$X_{(n)}-u$ over a threshold $u$. In order
for this approach to make sense in the
study of the intensities of solar flares, it is
necessary that $X_{(n)}$  has a distribution
with a tail similar to a power law tail, as
depicted in Equation~\ref{ec1}.
In this case, it is expected that the excesses
$X_{(n)}-u$ has a Pareto
distribution defined by
\begin{eqnarray}\label{e5}
H(x)=1-\Big(1+\frac{\kern.03cmx}
{\sigma}\Big)^{-\alpha},
\end{eqnarray}
where $\sigma$ is a scale parameter
and $\alpha$ is the shape parameter
we are interested in. Since $X$ and
$X_{(n)}$ belong to the same domain 
of attraction of a max-stable distribution
(in this case the Fr\'echet distribution),
then the value of $\alpha$ 
ought not to depend of $n$, and we
have departed from the custom of taking
$n=1$ and instead have applied the method
with $n=278$. This departure from
the usual way of applying the threshold
approach was motivated by the instability
of the results obtained when taking
$n=1$. In Subsection~\ref{s3.6}
we will suggest a reason 
for the instability of the estimation
of parameters when taking $n=1$.

We applied the threshold approach with
a block size given by
$n=278$ and with $u=9\times10^{-5}$ as
the chosen threshold.
With these values for $n$ and $u$, we obtained
a set of excesses of size 153.
Then we used the
maximum likelihood method for the
estimation of the parameters
in the generalized Pareto distribution
given by Equation~\ref{e5}. For these parameters
we obtained $\hat{\sigma}=7.59\times10^{-4}$
and $\hat{\alpha}=3.34$.
For the null hypothesis
stating that the 153 excesses
come from a Pareto distribution
given by Equation~\ref{e5},
the Pearson chi squared statistic has
a $p$-value of 0.875. In Table~\ref{t3}
we report a 95\% confidence interval
for the parameter $\alpha$. This
confidence interval was computed by
the profile likelihood method.

\subsection{Combining a log-logistic
and a Pareto parts}
\label{s3.4}

\begin{figure}
\begin{center}
\includegraphics[width=12cm]{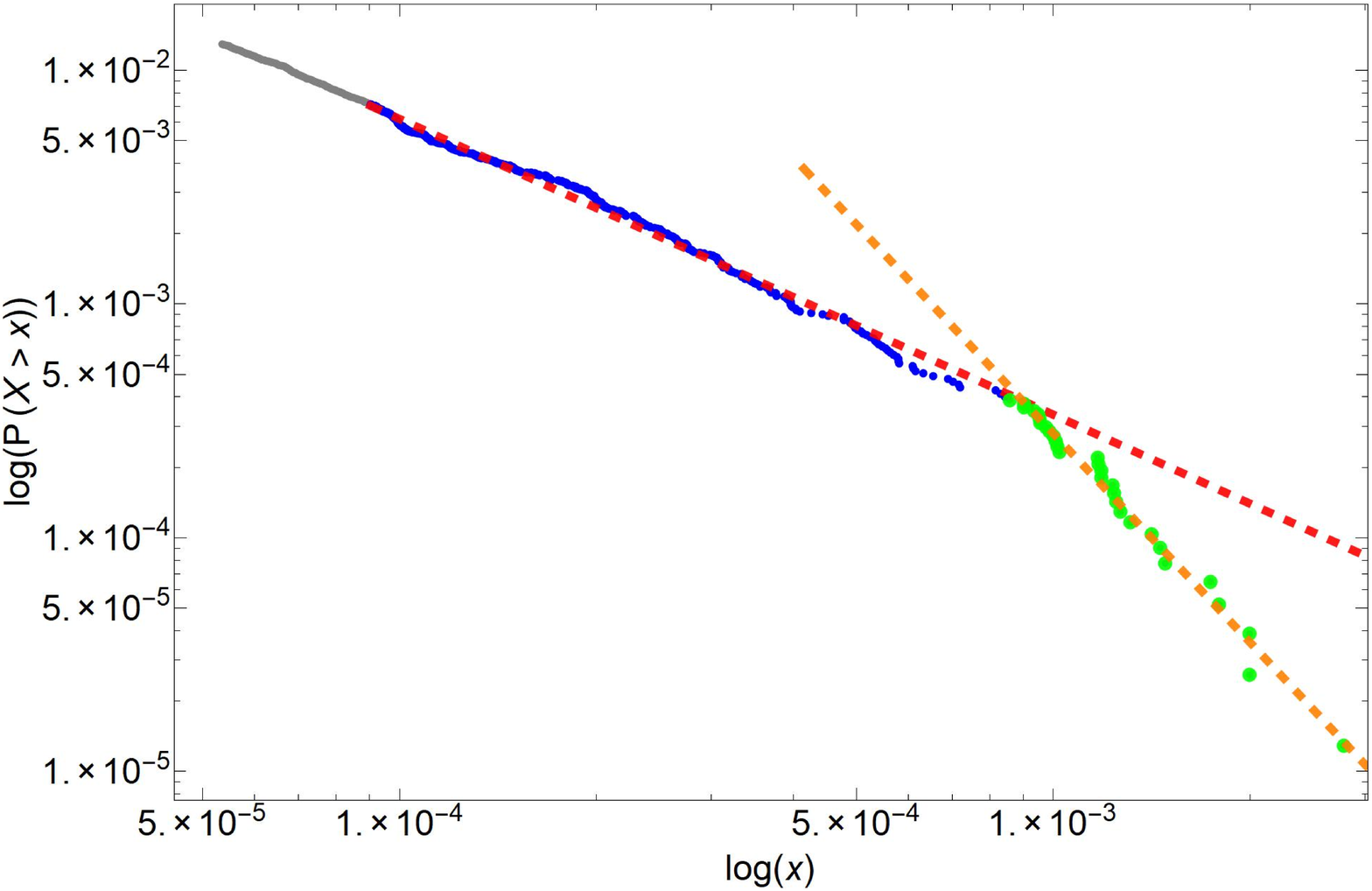}
\caption{The log-log plot for the 1\,000 most
extreme points in the data set. This figure
shows that a change in the slope of the fitted
\textit{straight line segments} is taking place as
we move from \textit{left to right}.}
\label{fig5}
\end{center}
\end{figure}

\begin{figure}
\begin{center}
\includegraphics[width=10cm]{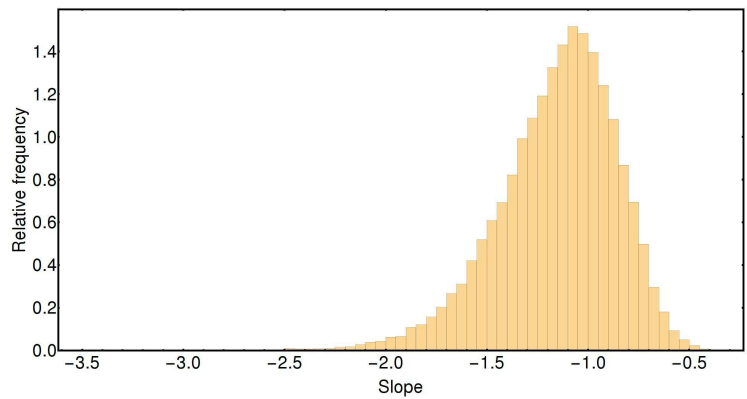}
\caption{The distribution of
the slope of the fitted regression 
line to the set of the 28 top extremes
in $10^4$ synthetic samples from a
log-logistic distribution with
$\alpha=1.27$. From this histogram,
it follows that the event $\{
\hat{\alpha}>3
\}$ has a small probability of
occurrence. Here $\hat{\alpha}$ is
computed from the 28 most extreme
data points in a sample.}
\label{fig8}
\end{center}
\end{figure}

In Figure~\ref{fig5} we present
the log-log plot of the 1\,000 most
extreme values in the data set under
study. In the horizontal axis of this
Figure~\ref{fig5} it is plotted
the quantity $\log x$, and
in the vertical axis the quantity
$\log\P\big\{X>x\big\}$, where
$X$ represents an observed solar flare intensity
in our data set. In this figure,
the points corresponding to
intensities $X$ lying in the
interval $(9\times10^{-5},8.5\times10^{-4})$
are plotted in blue color. The number of
solar flare intensities $X$ lying in
this interval equals 530. Drawn in
red color, in Figure~\ref{fig5} we can
see the straight line fitted by least
squares to the color blue points. 
The slope and intercept of
the fitted line equal $-1.2689$ and
$-16.772$ respectively. Henceforth, 
on a first stance,
one might expect that when
$x$ is large (larger than
$9\times10^{-5}$ W m$^{-2}$), then we have the
estimate
\begin{eqnarray}\label{e6}
\P\big\{X>x\big\}\>=\>
\frac{5.1991\times10^{-8}}
{x^{1.2689}}.
\end{eqnarray}
Also in Figure~\ref{fig5}, we have plotted
in green color the points corresponding to
the 28 largest values of $X$, that is
to say, those points for which 
$X>8.5\times10^{-4}$ W m$^{-2}$. 
Drawn in
orange color, in Figure~\ref{fig5} we can
see the straight line that
best fits the color green points. 
The slope and intercept of
the fitted line equal $-3.0382$ and
$-29.147$ respectively. Thus, when
$x$ is large (larger than
$8.5\times10^{-4}$), then we have
\begin{eqnarray}\label{e7}
\P\big\{X>x\big\}\>=\>
\frac{2.1954\times10^{-13}}
{x^{3.0382}}.
\end{eqnarray}
The question now arises as to whether
the difference in the two exponents
of $x$ in Equations~\ref{e6} 
and~\ref{e7} is due to a pure
random fluctuation or whether there
is a physical cause that explains
this difference. Now let us consider
the null hypothesis ${\rm H}_0$ stating that
the difference in the exponents of $x$ in
equations~\ref{e6} and~\ref{e7} is
explained as a random fluctuation.
In order to test ${\rm H}_0$, we produced a total
of $10^{4}$ synthetic samples,
each of size 77\,370, from a population
with log-logistic distribution with
parameters $\alpha=1.27$ and
$\sigma=1.7\times10^{-6}$. Here, the
value of $\alpha$ was chosen to be
approximately the same as in Equation~\ref{e6}
and the value of $\sigma$ was selected by
maximum likelihood form a sub-sample
of size 1\,000 of the solar flare
intensities data set. A comparison of
Equations~\ref{e2} and~\ref{e6} makes
it clear that the tail of a
log-logistic distribution will have
a tail as depicted by Equation~\ref{e6}.
See Figure~\ref{fig8} where we present
the histogram 
depicting the distribution
the slope of the fitted straight line
to the 28 most extreme values
in each of the above 
$10^{4}$ synthetic samples.
From these synthetic samples,
we estimated the probability
that the 28 most extreme values
in one of these samples have
a distribution as given in
Equation~\ref{e7} with an exponent
$\alpha$ larger in
absolute value than or equal to $3.0382$.
This probability 
is approximately $4\times10^{-5}$. 
These Monte Carlo computations lead
us to reject ${\rm H}_0$, that is to say, 
we conclude that the difference in the two
exponents of $x$ in Equations~\ref{e6} 
and~\ref{e7} is not explained as a result
of a pure random fluctuation.

The rejection of ${\rm H}_0$ suggests 
that an attenuation phenomenon
is taking place that results in that the
most extreme solar flares (those larger than
$8.5\times10^{-4}$) have a lesser probability
of occurrence than what is
predicted from the model given
by Equation~\ref{e6}. 

In view of the
former considerations we now propose to model the
data set under study by combining a log-logistic
distribution and a Pareto distribution. 
In order to state with
precision this idea, let us consider two
probability distributions $F_1$ and $F_2$. We
write $\overline{F}(x)=1-F(x)$ for the 
complementary cumulative distribution function.
Let us say that $G$ is the composition 
at $\sigma_2$ of $F_1$ and $F_2$, if
\begin{eqnarray}\label{e8}
G(x)=\cases{
F_1(x) & if $x\leq\sigma_2$,\cr
F_1(\sigma_2)+\overline{F}_1(\sigma_2)F_2(x)
& if $x>\sigma_2$.}
\end{eqnarray}
Notice that if $x>\sigma_2$ then
$\overline{G}(x)=\overline{F}_1(\sigma_2)
\overline{F}_2(x)$. 
Notice also
that if $G(x)$ is defined as in Equation~\ref{e8},
then it is continuous at $x=\sigma_2$ and $G(x)\to1$
as $x\to\infty$. This last condition must be
satisfied by every cumulative 
probability distribution function.
In the special case that $F_2$ is
a Pareto distribution of the form
\begin{eqnarray}\label{e9}
F_2(x)=1-\Big(\frac{x}{\sigma_2}
\Big)^{-\alpha_2}
\end{eqnarray}
for $x\geq\sigma_2$, then the
representation of $G$ in log-log
scale is given by
\[
\log \overline{G}(e^x)=
\cases{
\log \overline{F_1}(e^x)
& if $x\leq\sigma_2$,\cr
-\alpha_2x+\alpha_2\log\sigma_2+
\log \overline{F_1}(e^{\sigma_2})
& if $x>\sigma_2$.}
\]
Now we let $F_1$ be a log-logistic
distribution with parameters $\alpha_1$ and
$\sigma_1$ and $\overline{F}_2(x)=
(\sigma_2/x)^{\alpha_2}$
the Pareto distribution with
parameters $\alpha_2$ and $\sigma_2$.
We let $G_1$ be
the composition at $\sigma_2$ of these
$F_1$ and $F_2$. We propose to model
the data set under study by this
distribution $G_1$.

The distribution $G_1$ is determined
by its four parameters $\alpha_1$,
$\sigma_1$, $\alpha_2$ and $\sigma_2$.
For the estimation of these parameters,
we divided the data set into two subsets
$S_1$ and $S_2$. In $S_1$ we included
all data points of magnitude less than or
equal to $\sigma_2$, where $\sigma_2$ is assumed
to be known. In $S_2$ we included all data
points of magnitude greater than $\sigma_2$.
Given a numerical value $\hat{\sigma}_2$ for $\sigma_2$,
the first two parameters $\alpha_1$ and
$\sigma_1$ where estimated by the method
of moments from the data points in $S_1$.
Here we are assuming that $S_1$ is a
sample from a log-logistic distribution
truncated at $\sigma_2$.
With the same given value for $\sigma_2$,
the value of $\alpha_2$ was estimated
by the maximum likelihood method. Once
we have numerical values for $\hat{\alpha}_1$,
$\hat{\sigma}_1$,  $\hat{\alpha}_2$ and
$\hat{\sigma}_2$, the average distance
$\bar{D}(\hat{\sigma}_2)$ from the distribution
$\hat{G}_1$ to the empirical distribution
was computed. By letting $\hat{\sigma}_2$ assume
a whole range of values, we found that
$\bar{D}(\hat{\sigma}_2)$ assumes a minimum value
of 0.0139
when we set $\hat{\sigma}_2=9\times10^{-4}$.
It is interesting
to note that with the above optimal
value for $\hat{\alpha}_2$ the
set $S_2$ contains 28 points. This
vindicates the more naive approach
for the determination of the
stochastic behavior of the few most
extreme values in our data set
presented at the beginning of
this subsection.

Given this optimal value
$\hat{\sigma}_2=9\times10^{-4}$,
the rest of the parameters of $G_1$ were
determined as stated above. For the
first parameter we found that
$\hat{\alpha}_1=1.2409$ with 
(1.19, 1.30)
as 95\% confidence interval.
For the second parameter we found that
$\hat{\sigma}_1=0.0160$ with
(0.015, 0.017)
as 95\% confidence interval. For
the third parameter of $G_1$ we found that
$\hat{\alpha}_2=3.04$ with
(2.29, 4.33)
as 95\% confidence interval. 
All these are bootstrap 
confidence intervals computed
by re-sampling the original data set
$10^4$ times.

It is interesting to compare the
upper limit of the above confidence
interval for $\alpha_1$ with
the lower limit of the confidence
interval for $\alpha_2$. The difference
of these is $2.29-1.30=0.99$. This
positive difference is in agreement 
with the rejection of ${\rm H}_0$ above,
that is to say, the rejection of the
claim that the few (about 28) most
extreme values in our data set
follow a probability law with a
tail behavior as depicted in
Equation~\ref{e6}.

In Figure~\ref{fig6} we present
the log-log plot of the solar flare intensities
together with the log-log plot of the
complementary distribution
function $\overline{G}_1(x)$, where $G_1(x)$ is as
just determined (we omit the hat
symbol).

\subsection{A log-logistic part
and a tempered Pareto part}
\label{s3.5}

Although with the composition of a
log-logistic and a Pareto distributions
we have achieved a reasonably good fit to the
data under study, it should be
remarked that
the distribution $G_1$ considered in 
Subsection~\ref{s3.4} is not
without inconveniences. Indeed, $G_1$
is not differentiable at $\sigma_2$
and it is questionable that a non
differentiable distribution is to
be acceptable as a model for our
data set. On the other hand, the
number of data points explained by
the Pareto part of $G_1$ is only 28, and this
is a small number. For the
purpose of the estimation of
parameters of a proposed model, it is desirable
to work with as large a sample
as it is possible. In this subsection
we consider as a model for our
data set a composition of a
log-logistic distribution and
a tempered Pareto distribution
given by
\begin{eqnarray}\label{10}
\overline{T}(x)=c\kern.03cm
x^{-\alpha_2}e^{-\beta x}
\end{eqnarray}
when $x>\sigma_2$ and $\overline{T}(x)=0$
when $x\leq\sigma_2$.
Here $c=\sigma_2^{\alpha_2} e^{\beta\sigma_2}$
is the corresponding normalizing constant and
$\sigma_2$, $\alpha_2$ and $\beta$ are
parameters of the tempered Pareto
distribution. See \cite{Meerschaert} and
\cite{Clauset} for information about
the tempered Pareto distribution.
Now we consider the probability 
distribution function $G_2$ that results
from the composition at $\sigma_2$
of a log-logistic
distribution with parameters $\alpha_1$,
$\sigma_1$ (Equation~\ref{e2}) 
and a tempered Pareto distribution
with parameters $\alpha_1$, $\beta$,
$\sigma_2$ (Equation~\ref{10}).

For the estimation of the five parameters
of $G_2$, we proceeded as with the estimation
of the four parameters of $G_1$. Thus, we
considered a set $S_1$ containing all data points
less than or equal to 
$\sigma_2$ and a set $S_2$ containing all
data points greater than $\sigma_2$. Given
a tentative numerical value $\hat{\sigma}_2$
for $\sigma_2$, we used the method of
moments in order to 
obtain estimates $\hat{\alpha}_1$
and $\hat{\sigma}_1$, determining the 
log-logistic part of $G_2$, from 
the set $S_1$. With this same numerical
value $\hat{\sigma}_2$, we used the data
points in $S_2$ and the method of
moments in order to obtain estimates
$\hat{\alpha}_2$ and $\hat{\beta}$ for the
remaining two parameters of the tempered 
Pareto part of $G_2$.

The average distance $\bar{D}(\hat{\sigma}_2)$ 
from the estimated distribution $\hat{G}_2$ and
the empirical distribution was computed for
a range of numerical values $\hat{\sigma}_2$.
We found that $\bar{D}(\hat{\sigma}_2)=
0.00777$ is the smallest value for this
distance, and this smallest value is
achieved when $\hat{\sigma}_2=8\times10^{-6}$.
With this optimal value for $\hat{\sigma}_2$
we found the estimates for the remaining
parameters of $G_2$. For the
first parameter we found that
$\hat{\alpha}_1=1.358$ with 
(1.338, 1.377)
as 95\% confidence interval.
For the second parameter we found that
$\hat{\sigma}_1=0.0172$ with
(0.0169, 0.0174)
as 95\% confidence interval. For
the third parameter of $G_2$ we found that
$\hat{\alpha}_2=1.106$ with
(1.074, 1.136)
as 95\% confidence interval. 
For the fourth parameter of $G_2$ we found that
$\hat{\beta}=0.098$ with
(0.073, 0.138)
as 95\% confidence interval. 
All the these are bootstrap confidence
intervals.

In Figure~\ref{fig7} we present
the log-log plot of the solar flare intensities
together with the log-log plot of the
complementary distribution function 
$\overline{G}_2(x)$, where $G_2(x)$ is as
just determined (with the hat
symbol omitted).

\subsection{Performance of the estimation procedures}
\label{s3.6}
 
In this subsection we reflect on the
consequences for the estimation of
the parameter $\alpha$ by applying the
block and threshold methods 
to a sample that presents the 
same dual tail behaviour as does
the data set under study, and
which was put in evidence in
Subsection~\ref{s3.4}.

When applying the block approach to
the estimation of $\alpha$, a set
$S_{59}$ of 59 maxima was formed and
then it was used for the
estimation by maximum likelihood of the 
parameters of the ${\rm G}_{\rm ev}$
distribution given by Equation~\ref{e4},
where $\alpha$ is given by $1/\xi$.
If the number of blocks is large
(larger than 28), then it
is highly likely that this sample
$S_{59}$ comes from a distribution
that results from a mixture of the two
distributions given by Equation~\ref{e7}
and a truncated version of the
distribution determined by 
Equation~\ref{e6}\footnote{In terms of 
Figure~\ref{fig6}, the set $S_{59}$ would
be a combination of data points on
the left and on the right of the
dashed vertical line in the figure.}.
However, the fact that this sample
$S_{59}$  comes from a mixture
of distinct distributions is not taken
into account by the estimation
procedure of the parameters of the
${\rm G}_{\rm ev}$ distribution.
For example, when considering synthetic
samples from the distribution $G_1$
considered in Subsection~\ref{s3.4},
where $G_1$ is composed of two
distributions (log-logistic and
Pareto) having tail indices
$\alpha_1=1.27$ and $\alpha_2=3$,
then the estimate $\hat{\alpha}$
in the block approach will assume a value
in between $\alpha_1$ and $\alpha_2$,
unless the number of blocks is small,
in which case $\hat{\alpha}$ will be
closer to $\alpha_2$, but in this
case the estimation errors will be large.

On the other hand, when applying 
the threshold approach
to the solar flare data set, we
have a set $S_{28}$ of about 28
data points to which a straight line
with a slope about $-3$ can be fitted. From
this sample $S_{28}$ we are to estimate
the parameters of the distribution $H$
given in Equation~\ref{e5}. Yet, it is 
desirable to perform the estimation
procedure from a sample $S^*$ as large as
possible. One can obtain a larger sample
$S^*$ by choosing a lower threshold $u$.
However, if we choose a low enough 
threshold $u$, then our sample $S^*$
will come from a mixture of two
Pareto distributions, where one of
them has been truncated. When plotted
on log-log scale, this sample $S^*$ will
align itself along two straight line
segments that meet in a corner, drawn
in gray color in parts (a) and (b) of
Figure~\ref{fig11}. In this Figure~\ref{fig11}
we have also drawn in red color the function
\[
\log H(e^x)=-\alpha\kern.03cmx+
\alpha\log\sigma-\alpha\log
\Big(1+\frac{\sigma}{e^x}\Big),
\] 
which is the representation in log-log
scale of the distribution $H$ given in
Equation~\ref{e5}. In this figure,
the function $\log H(e^x)$ has been drawn
so that it reproduces  the two slopes
of the gray broken line at the two points
drawn in blue color. Notice that when
these two blue points are sufficiently close
together, then the function  $\log H(e^x)$
has a large curvature that results  in an
overestimation of the parameter
$\alpha$, which in the figure corresponds
to the absolute value of the slope of the
right hand gray straight line segment.

\section{Return periods and expected
frequencies}\label{returnperiods}

In this section we report the
return levels corresponding to
distinct time periods which
result from the block, threshold,
log - logistic/Pareto and
log - logistic/tempered Pareto
approaches to the estimation of the tail 
behavior of the solar flare
intensities data set.

In Table~\ref{t4} we display
the return levels that correspond
to distinct return periods computed
by the block approach. For example,
for a return period of 10 years,
we expect to observe once, an
event of magnitude 
$1.68\times10^{-3}$ 
(W m$^{-2}$) or larger.
In the third and
fourth columns of this table, we
report the lower and upper bounds
of a 95\% confidence interval
for the return level. These confidence
intervals were computed by the
normal approximation method.

In Table~\ref{t5} we present
return levels together with their 95\%
confidence intervals that
correspond to distinct return
periods computed by the threshold approach.
These confidence intervals were
also computed by assuming that
we have a normal approximation
for the distribution of the
estimated parameters of the model.

In Table~\ref{t6} we present the
return levels together with their 95\%
confidence intervals that
correspond to distinct return
periods computed from the model
which results from the composition
of a log-logistic and a Pareto distributions
as discussed in Subsection~\ref{s3.4}.
In this case, the confidence intervals were
computed by the bootstrap method by resampling
the original data set $10^4$ times.

In Table~\ref{t7} we present the
return levels together with their 95\%
confidence intervals 
corresponding to distinct return
periods computed from the model
which results from the composition
of a log-logistic and a 
tempered Pareto distributions
as discussed in subsection~\ref{s3.5}.
These confidence intervals were also
computed by the bootstrap method.

From Tables~\ref{t4} to~\ref{t7} it is
evident that the return levels corresponding
to distinct periods are larger when
computed from the block and threshold
approaches than when they are computed form
the log-logistic/Pareto and log-logistic/tempered
Pareto models. It is reasonable to contend
that this is so because of the inability of the
first two approaches to properly deal with  
a sample that results from a mixture of two
distributions with two distinct tails, one
heavier than the other. It is quite evident
that, in the block approach, the inclusion
of data points (in the set $S_{59}$) from
the first part of the tail of our data
set (a tail with $\alpha=1.35$ according
to the first estimate in Table~\ref{t3})
results in an underestimation of $\alpha$,
which in turn results in an overestimation
of the return levels. For the threshold
approach, it is more difficult to
present clear and convincing reasons in 
order to support our contend that it 
also overestimates the return levels.
Let it be enough to remark that the
data set under study only allowed us to
apply the threshold method for 
estimating $\alpha$ with a rather
small set of data points from
which we estimated the parameters
of distribution $H$ in Equation~\ref{e5}.
Therefore, the threshold method as
applied here, does not allow us to
describe a long tail (over at
least two orders of magnitude) as
in the case of a distribution without
a dual behavior of its tail (see
\cite{Stumpf}).

By performing numerous Monte Carlo
simulations with synthetic samples
resembling the original data set
under study\footnote{For the generation
of these synthetic samples, we used both,
$G_1$ and $G_2$ as the parent 
distribution.}, it became clear that
the methods of the extreme value
theory use the available information
(in the sets $S_{59}$ and $S^*$ of
Subsection~\ref{s3.6}) more efficiently
than the use made of the information
(the whole data set) available to the
more naive approaches we followed
by using compositions of a log-logistic
and Pareto parts. It turns out 
that the information available to
the composition models is so
large (recall that the size of the entire data
set is 77\,370) that it outweighs
the advantage of using the methods
of the extreme value theory. The advantage
in the amount of information 
for the composition models over the
amount of information in the block and
threshold approaches is put into
evidence when considering the length
of the confidence intervals for $\alpha$
reported in Table~\ref{t3}. Indeed, 
we have shorter confidence intervals in
the composition models than in
the block and threshold approaches.

In Table~\ref{t8} we report the 
expected number of events of distinct
magnitude to be observed in one solar cycle (11 years) 
computed from the models which result
from the composition of a log-logistic and a Pareto distributions and the composition of a log-logistic
and tempered Pareto distributions.
The expected number of occurrences of the
event $\{X > x\}$ in a time 
period of $n$ years is equal to $\e(Y)$, where 
$Y$ is a binomial random variable 
with probability of successes 
$p=\P\{X > x\}$ and
number of trials $\tilde{n}= 365.24n d$, 
where $d=5.05248$ is the average number of 
flares observed per day. Notice that this
expected number of observations $\e(Y)$
is a function of $d$, which is the
number of events per day, and recall that $d$
is not constant over time.
In this Table~\ref{t8}, we also display the
corresponding expected number of events 
reported in the NOAA web page\footnote{
www.swpc.noaa.gov/noaa-scales-explanation}. 
We find a reasonable good
agreement between the web page results 
and the results we obtained form the
two composition models. The large amount
of information available for the
two composition models also results
in small confidence intervals for the
return levels in Tables~\ref{t6} 
and~\ref{t7}.

\section{Conclusions}\label{conclusions}

\begin{figure}
\begin{center}
\includegraphics[width=12cm]{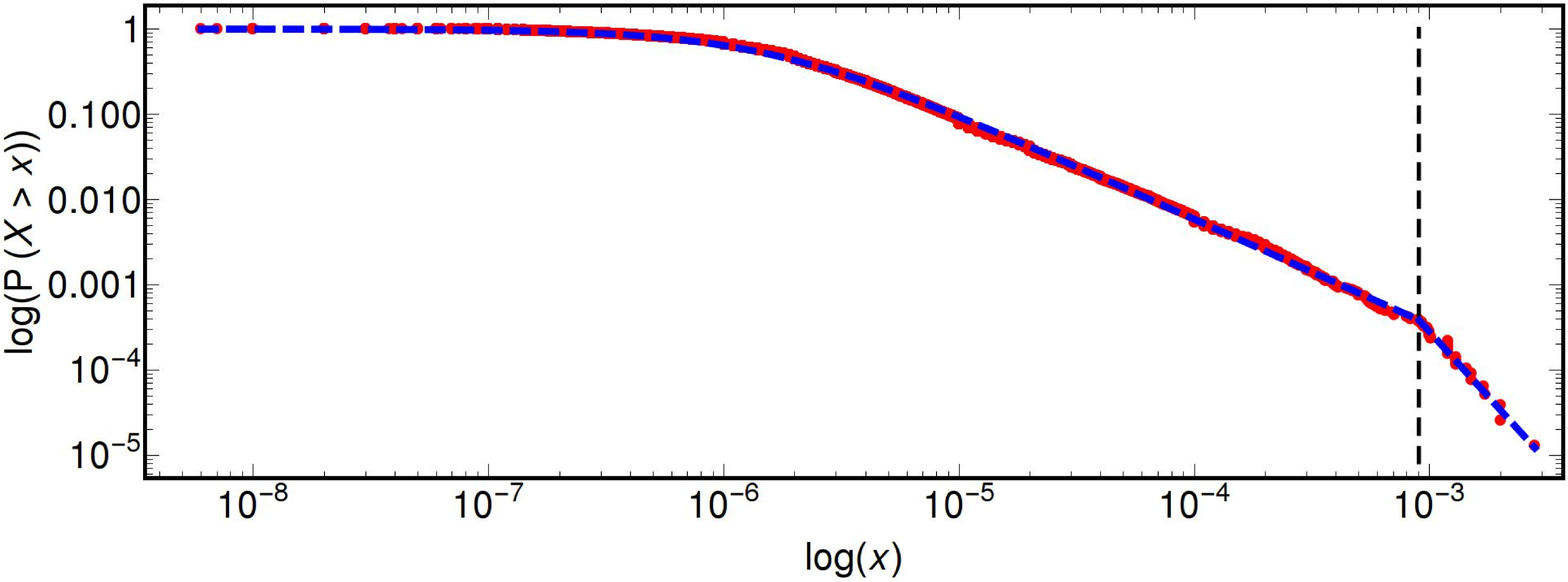}
\caption{The log-log plot for the full data set
and of $\overline{G}_1(x)$.  To the \textit{left} of
the \textit{dashed vertical line}, the log-logistic
distribution holds. To the \textit{right} of the
\textit{dashed line} the  
Pareto distribution holds.}
\label{fig6}
\end{center}
\end{figure}

\begin{figure}
\begin{center}
\includegraphics[width=12cm]{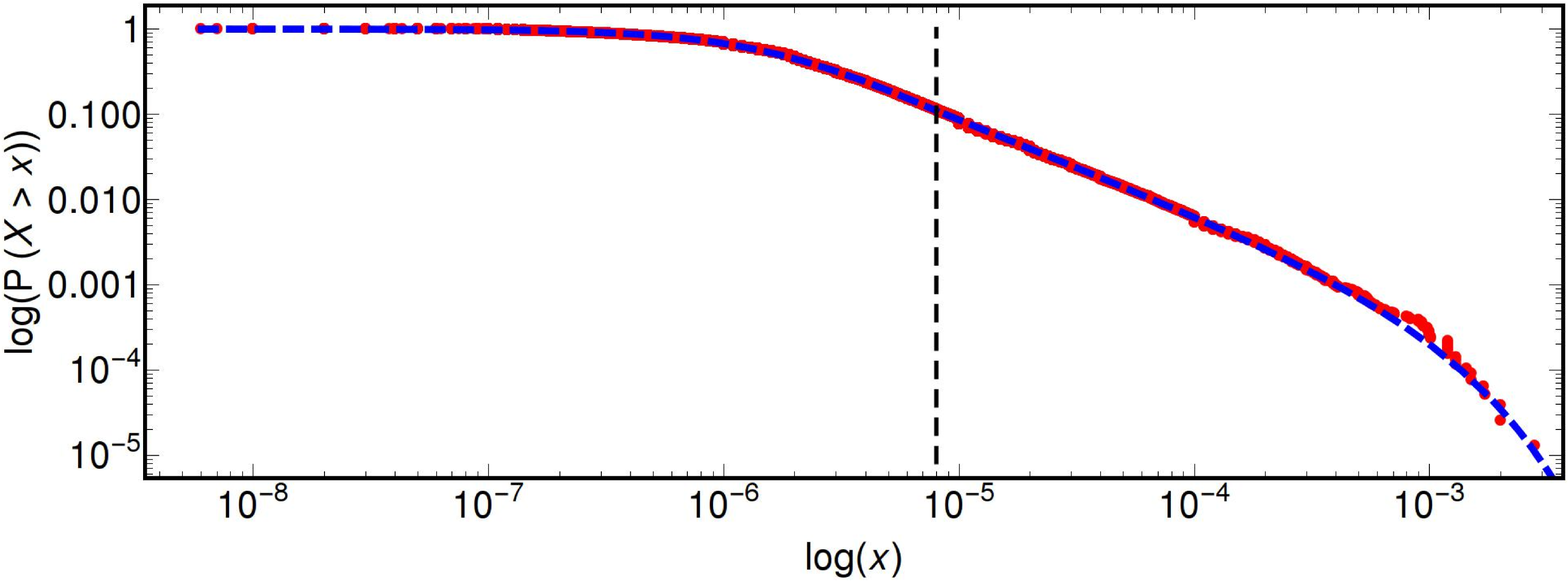}
\caption{The log-log plot for the full data set
and of $\overline{G}_2(x)$. To the \textit{left} of
the \textit{dashed vertical line}, the log-logistic
distribution holds. To the \textit{right} of the
\textit{dashed line} the tempered Pareto distribution holds.}
\label{fig7}
\end{center}
\end{figure}

\begin{figure}    
   \centerline{\hspace*{0.015\textwidth}
               \includegraphics[width=0.515\textwidth,clip=]{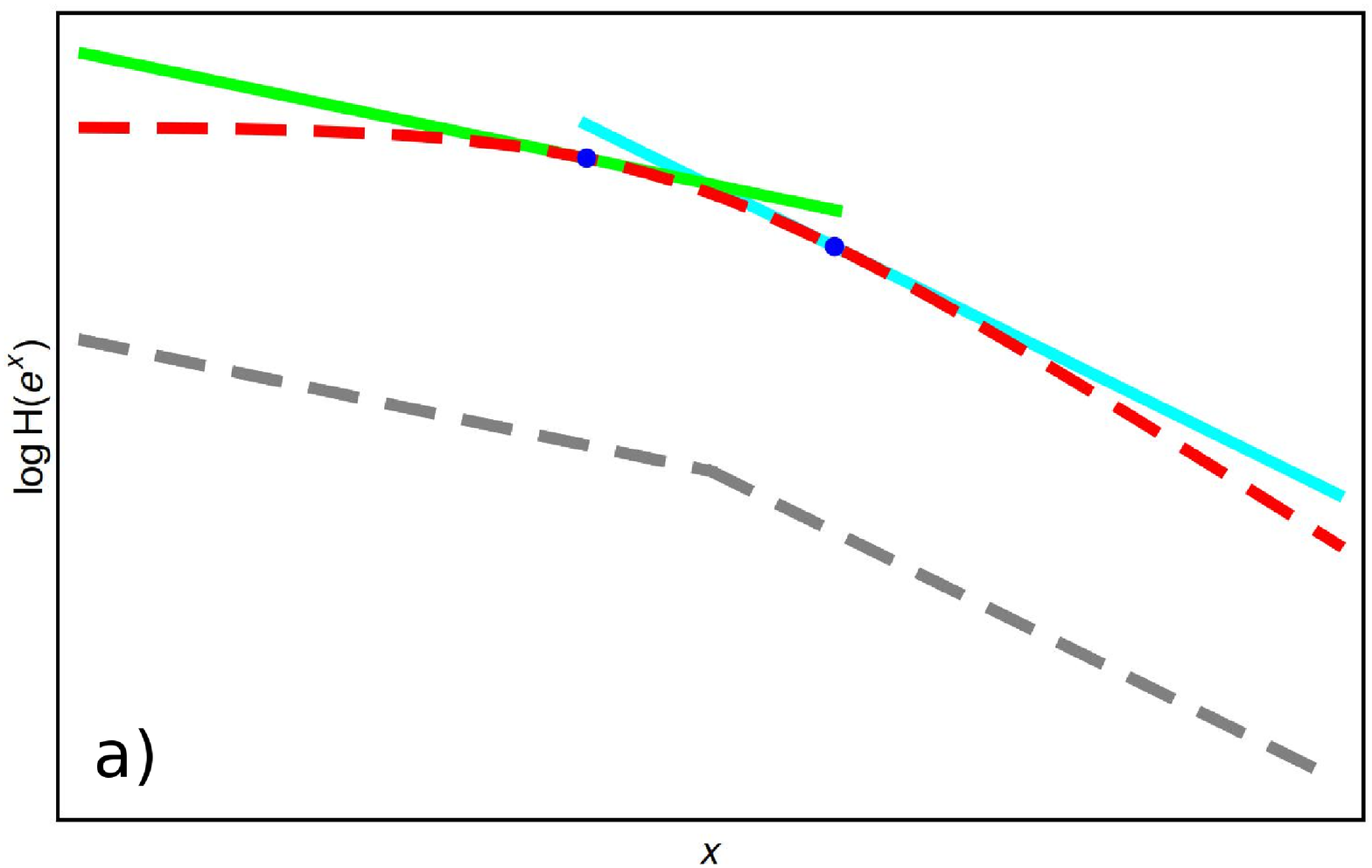}
               \hspace*{-0.03\textwidth}
               \includegraphics[width=0.515\textwidth,clip=]{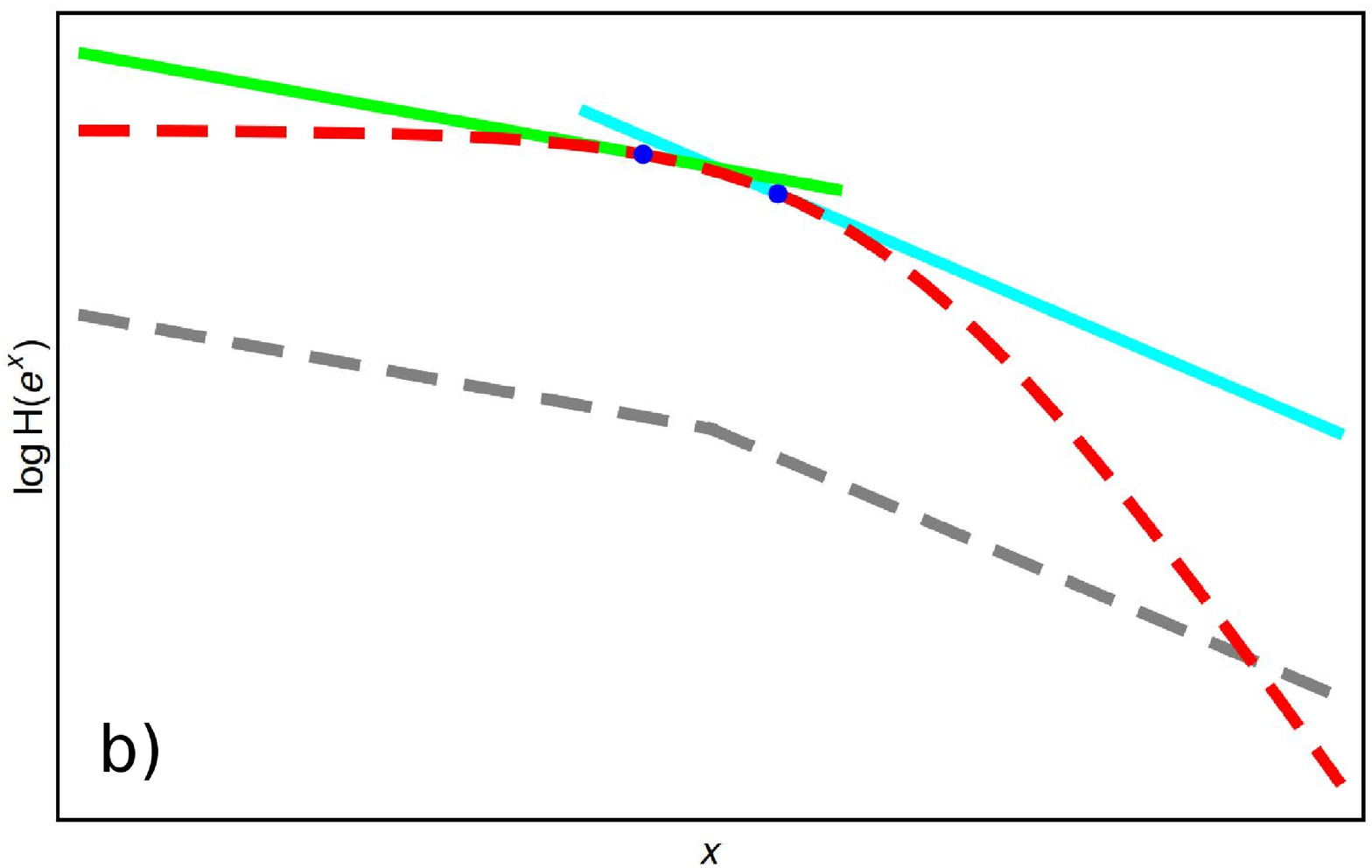}
              }
     \vspace{-0.35\textwidth}   
     \centerline{\Large \bf     
      \hspace{0.0 \textwidth}  \color{white}{(a)}
      \hspace{0.415\textwidth}  \color{white}{(b)}
         \hfill}
     \vspace{0.31\textwidth}    
              
\caption{Large curvature
        as misbehavior in the threshold approach.
        }
   \label{fig11}
   \end{figure}

The aim of this work at the beginning was
to determine the tail behavior of the
probability distribution of the
intensity $X$ of solar flares, as
depicted by the parameter $\alpha$
in Equation~\ref{ec1}. However, we latter found 
that the data set under study presents
a dual tail behavior. Indeed, we found
that the very most extreme values
in our data set are less intense than
what one would expect on account
of the behavior of the rest of
the data points. In terms of 
Figure~\ref{fig6}, the points on the
right of the dashed vertical line
behave differently from what one
would expect from the behavior of
the points on the left of the
dashed vertical line. This dual tail
behavior was confirmed by a hypothesis
test in Section~\ref{s3.4}. Now, it
is natural to pose the question of
whether the attenuation of the
intensities is a natural phenomenon
pertaining to the physics of the solar
activity or whether it is due to a 
threshold of the measuring 
instruments \citep{goes}.

It should be noted that
the dual behavior, or deviation from
a pure power law (as in 
Equation~\ref{ec1}), of the tail
of the distribution of the solar flare
intensities has already been noticed
in the literature (see \cite{Wheatland}
and \cite{Aschwanden}). In particular
\cite{Wheatland} applied
Bayesian methods in order to study
flare intensities in a single active
region, and found evidence for departure
from a pure power law behavior. Our findings
are in agreement with Wheatland's. 
The deviation from a pure power law
behavior in empirical data has been
observed, not only in studies pertaining
solar activity, but in other fields of
scientific inquiry as well (see \cite{Chinnery},
for example). As evidence mounts in favour to the
claim that pure power laws are not completely
suitable for describing extreme events in
natural phenomena, alternative models appear
in the literature 
(see \cite{Chakrabarty,Grabchak}) as 
modelling resources
whose usefulness ought to be explored in
the future.

Our results show that on average we can expect one solar flare
greater than X23
each 25 years, that is to say, one such 
event each two solar
cycles. The threshold of \~X20 is 
important because the energy 
level of saturation of the GOES 
spacecraft is X17.1
 \citep{goes}. 
After the flux of solar flares 
reaches X20 the instruments 
do not provide any more data. The peak 
of energy of the solar flare in 
progress remains unknown until 
forensic techniques allow us to 
determine the intensity of the 
flare (\cite{2005A&A...433.1133K}). 
However, it is mandatory to know 
the energy released in order 
to start the protocols of 
civil protection for Space Weather. 
Fortunately, events like 
Carrington's \citep{2013JSWSC...3A..31C} 
seem not to be very frequent (X40). 
Our results show that this 
kind of extreme events have a 
return period between 131 years 
(Log-Logistic/Pareto) and 238 
years (Log-logistic/Tempered Pareto).  

%

\begin{figure}
\begin{center}
\includegraphics[width=11.9cm]{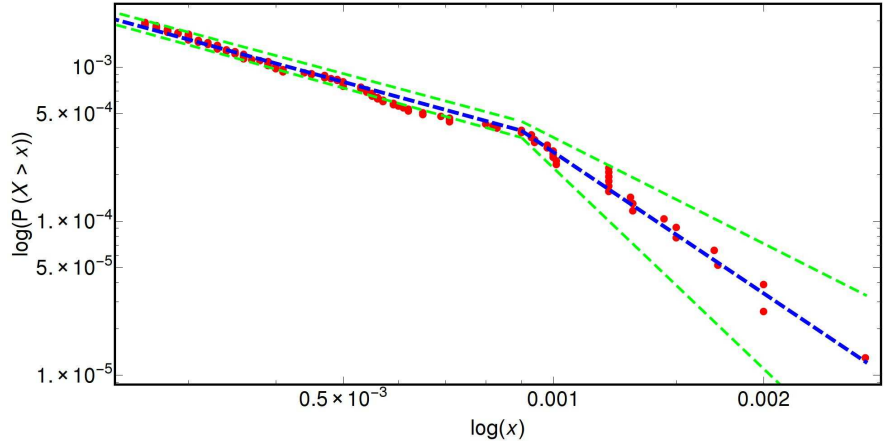}
\caption{This is the log-log plot of the 150 most
extreme points in the data set. The \textit{blue
dashed line} stands for the $\overline{G}_1$
complementary distribution 
function considered in Subsection~\ref{s3.4}.
The right hand part of the tails of
the two \textit{dashed green lines} have slopes
$-4.33$ and $-2.29$. These slopes correspond
to the 95\% confidence interval bounds reported
for the parameter $\alpha$ in the second part
of the log-logistic/Pareto model in Table~\ref{t3}.}
\label{fig12}
\end{center}
\end{figure}

\begin{table}
\caption{The distribution of
intensities in $1\times10^{-4}$ W m$^{-2}$
of the solar flares.}
\label{t1}
\begin{tabular}{|c | c | r | c | c | r |}
\hline
No. & Range & Data & No. & Range & Data\\ \hline\hline
1 & $< 0.05$ (C5) & 62\,364 &
11 & 8 (X8) to 10 (X10) & 11\\ \hline
2 & 0.05 (C5) to 0.1 (M1) & 8\,162 &
12 & 10 (X10) to 12 (X12)& 5\\ \hline
3 & 0.1 (M1) to 0.2 (M2) & 3\,635 &
13 & 12 (X12) to 14 (X14)& 9\\ \hline
4 & 0.2 (M2) to 0.3 (M3)& 1\,221 &
14 & 14 (X14) to 16 (X16)& 3\\ \hline
5 & 0.3 (M3) to 0.5 (M5) & 881 &
15 & 16 (X16) to 18 (X18)& 2\\ \hline
6 & 0.5 (M5) to 1 (X1) & 615 &
16 & 18 (X18) to 20 (X20)& 0\\ \hline
7 & 1 (X1) to 2 (X2) & 263 &
17 & 20 (X20) to 22 (X22)& 2\\ \hline
8 & 2 (X2) to 4 (X4) & 151 &
18 & 22 (X22) to 24 (X24)& 0\\ \hline
9 & 4 (X4) to 6 (X6) & 35 &
19 & 24 (X24) to 26 (X26)& 0\\ \hline
10 & 6 (X6) to 8 (X8) & 10 &
20 & 26 (X26) to 28 (X28)& 1\\ \hline
\end{tabular}
\end{table}

\begin{table}
\caption{The list of the most intense events in the data set (magnitude in $1\times10^{-4}$ W m$^{-2}$).}
\label{t2}
\begin{tabular}{|c|c|c|c|c|c|}
\hline          
No. & Magnitude 
& Date   & No. & Magnitude
& Date\\ \hline\hline    
1 & 28 (X28) & 4 Nov 2003  & 16 & 12 (X12) & 6 Jun 1982\\ \hline  
2 & 20 (X20) & 2 Apr 2001  & 17 & 12 (X12) & 1 Jun 1982\\ \hline  
3 & 20 (X20) & 16 Aug 1989 & 18 & 10 (X10) & 20 May 1984\\ \hline 
4 & 17 (X17) & 28 Oct 2003 & 19 & 10 (X10) & 17 Dec 1982\\ \hline 
5 & 17 (X17) & 7 Sept 2005 & 20 & 10 (X10) & 20 Oct 2003\\ \hline 
6 & 15 (X15) & 6 Mar 1989  & 21 & 10 (X10) & 9 Jun 1991\\ \hline  
7 & 15 (X15) & 11 Jul 1978 & 22 & 10 (X10) & 25 Jan 1991\\ \hline 
8 & 14 (X14) & 15 Apr 2001 & 23 & 10 (X10) & 29 Sept 1989\\ \hline
9 & 13 (X13) & 19 Oct 1989 & 24 & 10 (X10) & 9 Jul 1982\\ \hline  
10 & 13 (X13) & 25 Apr 1984& 25 & 9 (X9) & 6 Nov 1997\\ \hline  
11 & 13 (X13) & 15 Dec 1982& 26 & 9 (X9) & 22 Mar 1991\\ \hline 
12 & 12 (X12) & 15 Jun 1991& 27 & 9 (X9) & 6 Sept 2017\\ \hline 
13 & 12 (X12) & 11 Jun 1991& 28 & 9 (X9) & 24 May 1990\\ \hline 
14 & 12 (X12) & 6 Jun 1991 & 29 & 9 (X9) & 5 Dec 2006\\ \hline  
15 & 12 (X12) & 1 Jun 1991 & 30 & 9 (X9) & 6 Nov 1980\\ \hline  
\end{tabular}
\end{table}

\begin{table}
\caption{The estimated parameter $\alpha$ in various
models. In the column ``Fit" we report the 
Kolmogorov-Smirnov distance from the empirical
distribution to the distribution of the model
in turn. For the Pearson chi squared
statistic, the null hypothesis that the data 
fits the model is not rejected for those entries
in column ``$\alpha$"
marked with an asterisk.}
\label{t3}
\begin{tabular}{|r|c|c|r|c|}
\hline
Model & $\alpha$ & Conf. int. & Sample & Fit\\ \hline\hline
Log-logistic & $1.35^*$ & (1.25, 1.45) & 500 & 0.032\\ \hline
Block & $2.41^*$ & (1.54, 5.57)& 59 & 0.079\\ \hline
Threshold & $3.34^*$ & (1.83, 8.77) & 153 & 0.048\\ \hline
Log-logistic/Pareto 1 & 1.24 & (1.19, 1.30) & 
77\,342 & 0.055 \\ \hline
Log-logistic/Pareto 2 & $3.04^*$ & (2.29, 4.33) & 28 & 
$6\times10^{-5}$\\ \hline
Log-logistic/tempered Pareto 1 & 1.36 & (1.34, 1.38) &
68\,724 & 0.029\\ \hline
Log-logistic/tempered Pareto 2 & 1.11 & (1.07, 1.14) &
8\,646 & 0.01\\ \hline
\end{tabular}
\end{table}

\begin{table}
\caption{Return levels $\times
10^{-4}$ (W m$^{-2}$) for the block approach.}
\label{t4}
\begin{tabular}{|c|c|c|c|}
\hline
Period & Return level & Lower bound & Upper bound\\\hline
\hline
10 & 16.80 & 10.26 & 23.34\\\hline
25 & 26.49 & 12.39 & 40.58\\\hline
50 & 36.61 & 12.69 & 60.53\\\hline
100 & 50.07 & 12.69 & 89.24\\\hline
150 & 59.92 & 8.32 & 111.5\\\hline
\end{tabular}
\end{table}

\begin{table}
\caption{Return levels $\times
10^{-4}$ (W m$^{-2}$)
for the threshold approach.}
\label{t5}
\begin{tabular}{|c|c|c|c|}
\hline
Period & Return level & Lower bound & Upper bound\\\hline
\hline
10 & 19.11 & 12.10 & 26.12\\\hline
25 & 27.55 & 13.75 & 41.36\\\hline
50 & 35.67 & 14.00 & 57.34\\\hline
100 & 45.66 & 12.89 & 78.44\\\hline
150 & 52.55 & 11.35 & 93.75\\\hline
\end{tabular}
\end{table}

\begin{table}
\caption{Return levels $\times
10^{-4}$ (W m$^{-2}$)
for the log-logistic/Pareto model.}
\label{t6}
\begin{tabular}{|c|c|c|c|}
\hline
Period & Return level & Lower bound & 
Upper bound \\\hline
\hline
10 & 17.16 & 13.98 & 21.55 \\\hline
25 & 23.19 & 17.35 & 32.02 \\\hline
50 & 29.12 & 20.38 & 43.40 \\\hline
100 & 36.57 & 23.95 & 58.65 \\\hline
150 & 41.79 & 26.33 & 70.01 \\\hline
\end{tabular}
\end{table}

\begin{table}
\caption{Return levels $\times
10^{-4}$ (W m$^{-2}$)
for the log-logistic/tempered Pareto model.}
\label{t7}

\begin{tabular}{|c|c|c|c|}
\hline
Period & Return level & Lower bound & 
Upper bound\\\hline
\hline
10 & 17.19 & 14.04 & 20.37 \\\hline
25 & 23.17 & 18.51 & 27.94 \\\hline
50 & 28.07 & 22.15 & 34.27 \\\hline
100 & 33.24 & 25.95 & 41.00 \\\hline
150 & 36.36 & 28.21 & 45.04 \\\hline
\end{tabular}
\end{table}

\renewcommand{\arraystretch}{1}

\begin{table}
\caption{Expected frequency of event
occurrence computed by the
log-logistic/Pareto model (LL/P) and by
the log-logistic/tempered Pareto model (LL/TP).}
\label{t8}
\begin{tabular}{|c|c|c|c|c|c|}
\hline
Intensity (W/$\hbox{m}^2$)
& NOAA & LL/P& 
Conf. Int. & LL/TP & Conf. Int.\\\hline
\hline
$10^{-5}$ (M1)
& 2000 & 1894 & (1849, 1940) & 1745 & (1677, 1815)\\\hline
$5\times10^{-5}$ (M5)
& 350 & 279 & (254, 305) & 283 & (264, 303) \\\hline
$10^{-4}$ (X1)
& 175 & 119 & (104, 135) & 125 & (115, 136) \\\hline
$10^{-3}$ (X10)
& 8 & 7.87 & (6.04, 10.04) & 4.00 & (2.77, 5.24) \\\hline
$2\times10^{-3}$ (X20)
& $<1$ & 4.95 & (1.56, 9.27) & 0.70 & 
(0.33, 1.15)\\\hline
\end{tabular}
\end{table}

\begin{acks}
  The authors thank projects for Catedras Conacyt (Conacyt Fellow), Repositorios Institucionales (268273) and Ciencia Basica (254497).
\end{acks}



\clearpage
\newpage


\begin{thebibliography}{27}
\ifx\bisbn     \undefined \def\bisbn  #1{ISBN #1}\fi
\ifx\binits    \undefined \def\binits#1{#1}\fi
\ifx\bauthor   \undefined \def\bauthor#1{#1}\fi
\ifx\batitle   \undefined \def\batitle#1{#1}\fi
\ifx\bjtitle   \undefined \def\bjtitle#1{\textit{#1}}\fi
\ifx\bvolume   \undefined \def\bvolume#1{\textbf{#1}}\fi
\ifx\byear     \undefined \def\byear#1{#1}\fi
\ifx\bissue    \undefined \def\bissue#1{#1}\fi
\ifx\bfpage    \undefined \def\bfpage#1{#1}\fi
\ifx\blpage    \undefined \def\blpage #1{#1}\fi
\ifx\burl      \undefined \def\burl#1{\textsf{#1}}\fi
\ifx\href      \undefined \def\href#1#2{\textsf{#2}}\fi
\ifx\betal     \undefined \def\betal{\textit{et al.}}\fi
\ifx\bctitle   \undefined \def\bctitle#1{#1}\fi
\ifx\beditor   \undefined \def\beditor#1{#1}\fi
\ifx\bbtitle   \undefined \def\bbtitle#1{\textit{#1}}\fi
\ifx\bedition  \undefined \def\bedition#1{#1}\fi
\ifx\bseriesno \undefined \def\bseriesno#1{\textbf{#1}}\fi
\ifx\blocation \undefined \def\blocation#1{#1}\fi
\ifx\bsertitle \undefined \def\bsertitle#1{\textit{#1}}\fi
\ifx\bsnm      \undefined \def\bsnm#1{#1}\fi
\ifx\bsuffix   \undefined \def\bsuffix#1{#1}\fi
\ifx\bparticle \undefined \def\bparticle#1{#1}\fi
\ifx\barticle  \undefined \def\barticle#1{}\fi
\ifx\binstitute  \undefined \def\binstitute#1{#1}\fi
\ifx\bpublisher  \undefined \def\bpublisher#1{#1}\fi
\ifx\doiurl    \undefined
  \def\doiurl#1{\href{http://dx.doi.org/#1}{\textsf{DOI}}}\fi
\ifx\arxivurl  \undefined
  \def\arxivurl#1{\href{http://arxiv.org/abs/#1}{\textsf{arXiv}}}\fi
\ifx\adsurl    \undefined
  \def\adsurl#1{\href{http://adsabs.harvard.edu/abs/#1}{\textsf{ADS}}}\fi
\ifx\botherref \undefined \def\botherref#1{}\fi
\ifx\url       \undefined \def\url#1{\textsf{#1}}\fi
\ifx\bchapter  \undefined \def\bchapter#1{}\fi
\ifx\bbook     \undefined \def\bbook#1{}\fi
\ifx\bcomment  \undefined \def\bcomment#1{#1}\fi
\ifx\oauthor   \undefined \def\oauthor#1{#1}\fi
\ifx\citeauthoryear \undefined\def \citeauthoryear#1{#1}\fi
\ifx\endbibitem\undefined \def\endbibitem{}\fi
\ifx\bconflocation  \undefined \def\bconflocation#1{#1} \fi

\bibitem[\protect\citeauthoryear{{Ackermann}
  \textit{et~al.}}{2014}]{2014ApJ...787...15A}
\begin{barticle}
\bauthor{\bsnm{{Ackermann}}, \binits{M.}},
\bauthor{\bsnm{{Ajello}}, \binits{M.}},
\bauthor{\bsnm{{Albert}}, \binits{A.}},
\bauthor{\bsnm{{Allafort}}, \binits{A.}},
\bauthor{\bsnm{{Baldini}}, \binits{L.}},
\bauthor{\bsnm{{Barbiellini}}, \binits{G.}},
\bauthor{\bsnm{{Bastieri}}, \binits{D.}},
\bauthor{\bsnm{{Bechtol}}, \binits{K.}},
\bauthor{\bsnm{{Bellazzini}}, \binits{R.}},
\bauthor{\bsnm{{Bissaldi}}, \binits{E.}},
\bauthor{\bsnm{{Bonamente}}, \binits{E.}},
\bauthor{\bsnm{{Bottacini}}, \binits{E.}},
\bauthor{\bsnm{{Bouvier}}, \binits{A.}},
\bauthor{\bsnm{{Brandt}}, \binits{T.J.}},
\bauthor{\bsnm{{Bregeon}}, \binits{J.}},
\bauthor{\bsnm{{Brigida}}, \binits{M.}},
\bauthor{\bsnm{{Bruel}}, \binits{P.}},
\bauthor{\bsnm{{Buehler}}, \binits{R.}},
\bauthor{\bsnm{{Buson}}, \binits{S.}},
\bauthor{\bsnm{{Caliandro}}, \binits{G.A.}},
\bauthor{\bsnm{{Cameron}}, \binits{R.A.}},
\bauthor{\bsnm{{Caraveo}}, \binits{P.A.}},
\bauthor{\bsnm{{Cecchi}}, \binits{C.}},
\bauthor{\bsnm{{Charles}}, \binits{E.}},
\bauthor{\bsnm{{Chekhtman}}, \binits{A.}},
\bauthor{\bsnm{{Chen}}, \binits{Q.}},
\bauthor{\bsnm{{Chiang}}, \binits{J.}},
\bauthor{\bsnm{{Chiaro}}, \binits{G.}},
\bauthor{\bsnm{{Ciprini}}, \binits{S.}},
\bauthor{\bsnm{{Claus}}, \binits{R.}},
\bauthor{\bsnm{{Cohen-Tanugi}}, \binits{J.}},
\bauthor{\bsnm{{Conrad}}, \binits{J.}},
\bauthor{\bsnm{{Cutini}}, \binits{S.}},
\bauthor{\bsnm{{D'Ammando}}, \binits{F.}},
\bauthor{\bsnm{{de Angelis}}, \binits{A.}},
\bauthor{\bsnm{{de Palma}}, \binits{F.}},
\bauthor{\bsnm{{Dermer}}, \binits{C.D.}},
\bauthor{\bsnm{{Desiante}}, \binits{R.}},
\bauthor{\bsnm{{Digel}}, \binits{S.W.}},
\bauthor{\bsnm{{Di Venere}}, \binits{L.}},
\bauthor{\bsnm{{Silva}}, \binits{E.d.C.e.}},
\bauthor{\bsnm{{Drell}}, \binits{P.S.}},
\bauthor{\bsnm{{Drlica-Wagner}}, \binits{A.}},
\bauthor{\bsnm{{Favuzzi}}, \binits{C.}},
\bauthor{\bsnm{{Fegan}}, \binits{S.J.}},
\bauthor{\bsnm{{Focke}}, \binits{W.B.}},
\bauthor{\bsnm{{Franckowiak}}, \binits{A.}},
\bauthor{\bsnm{{Fukazawa}}, \binits{Y.}},
\bauthor{\bsnm{{Funk}}, \binits{S.}},
\bauthor{\bsnm{{Fusco}}, \binits{P.}},
\bauthor{\bsnm{{Gargano}}, \binits{F.}},
\bauthor{\bsnm{{Gasparrini}}, \binits{D.}},
\bauthor{\bsnm{{Germani}}, \binits{S.}},
\bauthor{\bsnm{{Giglietto}}, \binits{N.}},
\bauthor{\bsnm{{Giordano}}, \binits{F.}},
\bauthor{\bsnm{{Giroletti}}, \binits{M.}},
\bauthor{\bsnm{{Glanzman}}, \binits{T.}},
\bauthor{\bsnm{{Godfrey}}, \binits{G.}},
\bauthor{\bsnm{{Grenier}}, \binits{I.A.}},
\bauthor{\bsnm{{Grove}}, \binits{J.E.}},
\bauthor{\bsnm{{Guiriec}}, \binits{S.}},
\bauthor{\bsnm{{Hadasch}}, \binits{D.}},
\bauthor{\bsnm{{Hayashida}}, \binits{M.}},
\bauthor{\bsnm{{Hays}}, \binits{E.}},
\bauthor{\bsnm{{Horan}}, \binits{D.}},
\bauthor{\bsnm{{Hughes}}, \binits{R.E.}},
\bauthor{\bsnm{{Inoue}}, \binits{Y.}},
\bauthor{\bsnm{{Jackson}}, \binits{M.S.}},
\bauthor{\bsnm{{Jogler}}, \binits{T.}},
\bauthor{\bsnm{{J{\'o}hannesson}}, \binits{G.}},
\bauthor{\bsnm{{Johnson}}, \binits{W.N.}},
\bauthor{\bsnm{{Kamae}}, \binits{T.}},
\bauthor{\bsnm{{Kawano}}, \binits{T.}},
\bauthor{\bsnm{{Kn{\"o}dlseder}}, \binits{J.}},
\bauthor{\bsnm{{Kuss}}, \binits{M.}},
\bauthor{\bsnm{{Lande}}, \binits{J.}},
\bauthor{\bsnm{{Larsson}}, \binits{S.}},
\bauthor{\bsnm{{Latronico}}, \binits{L.}},
\bauthor{\bsnm{{Lemoine-Goumard}}, \binits{M.}},
\bauthor{\bsnm{{Longo}}, \binits{F.}},
\bauthor{\bsnm{{Loparco}}, \binits{F.}},
\bauthor{\bsnm{{Lott}}, \binits{B.}},
\bauthor{\bsnm{{Lovellette}}, \binits{M.N.}},
\bauthor{\bsnm{{Lubrano}}, \binits{P.}},
\bauthor{\bsnm{{Mayer}}, \binits{M.}},
\bauthor{\bsnm{{Mazziotta}}, \binits{M.N.}},
\bauthor{\bsnm{{McEnery}}, \binits{J.E.}},
\bauthor{\bsnm{{Michelson}}, \binits{P.F.}},
\bauthor{\bsnm{{Mizuno}}, \binits{T.}},
\bauthor{\bsnm{{Moiseev}}, \binits{A.A.}},
\bauthor{\bsnm{{Monte}}, \binits{C.}},
\bauthor{\bsnm{{Monzani}}, \binits{M.E.}},
\bauthor{\bsnm{{Moretti}}, \binits{E.}},
\bauthor{\bsnm{{Morselli}}, \binits{A.}},
\bauthor{\bsnm{{Moskalenko}}, \binits{I.V.}},
\bauthor{\bsnm{{Murgia}}, \binits{S.}},
\bauthor{\bsnm{{Murphy}}, \binits{R.}},
\bauthor{\bsnm{{Nemmen}}, \binits{R.}},
\bauthor{\bsnm{{Nuss}}, \binits{E.}},
\bauthor{\bsnm{{Ohno}}, \binits{M.}},
\bauthor{\bsnm{{Ohsugi}}, \binits{T.}},
\bauthor{\bsnm{{Okumura}}, \binits{A.}},
\bauthor{\bsnm{{Omodei}}, \binits{N.}},
\bauthor{\bsnm{{Orienti}}, \binits{M.}},
\bauthor{\bsnm{{Orlando}}, \binits{E.}},
\bauthor{\bsnm{{Ormes}}, \binits{J.F.}},
\bauthor{\bsnm{{Paneque}}, \binits{D.}},
\bauthor{\bsnm{{Panetta}}, \binits{J.H.}},
\bauthor{\bsnm{{Perkins}}, \binits{J.S.}},
\bauthor{\bsnm{{Pesce-Rollins}}, \binits{M.}},
\bauthor{\bsnm{{Petrosian}}, \binits{V.}},
\bauthor{\bsnm{{Piron}}, \binits{F.}},
\bauthor{\bsnm{{Pivato}}, \binits{G.}},
\bauthor{\bsnm{{Porter}}, \binits{T.A.}},
\bauthor{\bsnm{{Rain{\`o}}}, \binits{S.}},
\bauthor{\bsnm{{Rando}}, \binits{R.}},
\bauthor{\bsnm{{Razzano}}, \binits{M.}},
\bauthor{\bsnm{{Reimer}}, \binits{A.}},
\bauthor{\bsnm{{Reimer}}, \binits{O.}},
\bauthor{\bsnm{{Ritz}}, \binits{S.}},
\bauthor{\bsnm{{Schulz}}, \binits{A.}},
\bauthor{\bsnm{{Sgr{\`o}}}, \binits{C.}},
\bauthor{\bsnm{{Siskind}}, \binits{E.J.}},
\bauthor{\bsnm{{Spandre}}, \binits{G.}},
\bauthor{\bsnm{{Spinelli}}, \binits{P.}},
\bauthor{\bsnm{{Takahashi}}, \binits{H.}},
\bauthor{\bsnm{{Takeuchi}}, \binits{Y.}},
\bauthor{\bsnm{{Tanaka}}, \binits{Y.}},
\bauthor{\bsnm{{Thayer}}, \binits{J.G.}},
\bauthor{\bsnm{{Thayer}}, \binits{J.B.}},
\bauthor{\bsnm{{Thompson}}, \binits{D.J.}},
\bauthor{\bsnm{{Tibaldo}}, \binits{L.}},
\bauthor{\bsnm{{Tinivella}}, \binits{M.}},
\bauthor{\bsnm{{Tosti}}, \binits{G.}},
\bauthor{\bsnm{{Troja}}, \binits{E.}},
\bauthor{\bsnm{{Tronconi}}, \binits{V.}},
\bauthor{\bsnm{{Usher}}, \binits{T.L.}},
\bauthor{\bsnm{{Vandenbroucke}}, \binits{J.}},
\bauthor{\bsnm{{Vasileiou}}, \binits{V.}},
\bauthor{\bsnm{{Vianello}}, \binits{G.}},
\bauthor{\bsnm{{Vitale}}, \binits{V.}},
\bauthor{\bsnm{{Werner}}, \binits{M.}},
\bauthor{\bsnm{{Winer}}, \binits{B.L.}},
\bauthor{\bsnm{{Wood}}, \binits{D.L.}},
\bauthor{\bsnm{{Wood}}, \binits{K.S.}},
\bauthor{\bsnm{{Wood}}, \binits{M.}},
\bauthor{\bsnm{{Yang}}, \binits{Z.}},
\bauthor{\bsnm{{Fermi LAT Collaboration}}}:
\byear{2014},
\batitle{{High-energy Gamma-Ray Emission from Solar Flares: Summary of Fermi
  Large Area Telescope Detections and Analysis of Two M-class Flares}}.
\bjtitle{\apj}
\bvolume{787},
\bfpage{15}.
\doiurl{10.1088/0004-637X/787/1/15}.
\adsurl{2014ApJ...787...15A}.
\end{barticle}
\endbibitem

\bibitem[\protect\citeauthoryear{Aschwanden and Freeland}{2012}]{Aschwanden}
\begin{barticle}
\bauthor{\bsnm{Aschwanden}, \binits{M.J.}},
\bauthor{\bsnm{Freeland}, \binits{S.L.}}:
\byear{2012},
\batitle{Automated solar flare statistics in soft x-rays over 37 years of goes
  observations: the invariance of self-organized criticality during three solar
  cycles}.
\bjtitle{The Astrophysical Journal}
\bvolume{754}(\bissue{2}),
\bfpage{112}.
\end{barticle}
\endbibitem

\bibitem[\protect\citeauthoryear{{Benz} and
  {G{\"u}del}}{2010}]{2010ARA&A..48..241B}
\begin{barticle}
\bauthor{\bsnm{{Benz}}, \binits{A.O.}},
\bauthor{\bsnm{{G{\"u}del}}, \binits{M.}}:
\byear{2010},
\batitle{{Physical Processes in Magnetically Driven Flares on the Sun, Stars,
  and Young Stellar Objects}}.
\bjtitle{Annual Review of Astronomy and Astrophysics}
\bvolume{48},
\bfpage{241}.
\doiurl{10.1146/annurev-astro-082708-101757}.
\adsurl{2010ARA\%26A..48..241B}.
\end{barticle}
\endbibitem

\bibitem[\protect\citeauthoryear{{Canfield}
  \textit{et~al.}}{1993}]{1993ApJ...411..362C}
\begin{barticle}
\bauthor{\bsnm{{Canfield}}, \binits{R.C.}},
\bauthor{\bsnm{{de La Beaujardiere}}, \binits{J.-F.}},
\bauthor{\bsnm{{Fan}}, \binits{Y.}},
\bauthor{\bsnm{{Leka}}, \binits{K.D.}},
\bauthor{\bsnm{{McClymont}}, \binits{A.N.}},
\bauthor{\bsnm{{Metcalf}}, \binits{T.R.}},
\bauthor{\bsnm{{Mickey}}, \binits{D.L.}},
\bauthor{\bsnm{{Wuelser}}, \binits{J.-P.}},
\bauthor{\bsnm{{Lites}}, \binits{B.W.}}:
\byear{1993},
\batitle{{The morphology of flare phenomena, magnetic fields, and electric
  currents in active regions. I - Introduction and methods}}.
\bjtitle{\apj}
\bvolume{411},
\bfpage{362}.
\doiurl{10.1086/172836}.
\adsurl{1993ApJ...411..362C}.
\end{barticle}
\endbibitem

\bibitem[\protect\citeauthoryear{Castillo \textit{et~al.}}{2005}]{Castillo}
\begin{botherref}
\oauthor{\bsnm{Castillo}, \binits{E.}},
\oauthor{\bsnm{Hadi}, \binits{A.S.}},
\oauthor{\bsnm{Balakrishnan}, \binits{N.}},
\oauthor{\bsnm{Sarabia}, \binits{J.-M.}}:
2005,
Extreme value and related models with applications in engineering and science.
\end{botherref}
\endbibitem

\bibitem[\protect\citeauthoryear{Chakrabarty and
  Samorodnitsky}{2012}]{Chakrabarty}
\begin{barticle}
\bauthor{\bsnm{Chakrabarty}, \binits{A.}},
\bauthor{\bsnm{Samorodnitsky}, \binits{G.}}:
\byear{2012},
\batitle{Understanding heavy tails in a bounded world or, is a truncated heavy
  tail heavy or not?}
\bjtitle{Stochastic Models}
\bvolume{28}(\bissue{1}),
\bfpage{109}.
\end{barticle}
\endbibitem

\bibitem[\protect\citeauthoryear{{Chamberlin} \textit{et~al.}}{2009}]{goes}
\begin{bchapter}
\bauthor{\bsnm{{Chamberlin}}, \binits{P.C.}},
\bauthor{\bsnm{{Woods}}, \binits{T.N.}},
\bauthor{\bsnm{{Eparvier}}, \binits{F.G.}},
\bauthor{\bsnm{{Jones}}, \binits{A.R.}}:
\byear{2009},
\bctitle{{Next generation {X}-ray sensor ({XRS}) for the {NOAA GOES-R}
  satellite series}}.
In: \bbtitle{Solar Physics and Space Weather Instrumentation III},
\bsertitle{Proceedings of the SPIE}
\bseriesno{7438},
\bfpage{743802}.
\doiurl{10.1117/12.826807}.
\adsurl{2009SPIE.7438E..02C}.
\end{bchapter}
\endbibitem

\bibitem[\protect\citeauthoryear{{Chertok} and
  {Belov}}{2017}]{2017SoPh..292..144C}
\begin{barticle}
\bauthor{\bsnm{{Chertok}}, \binits{I.M.}},
\bauthor{\bsnm{{Belov}}, \binits{A.V.}}:
\byear{2017},
\batitle{{Long- and Mid-Term Variations of the Soft X-ray Flare Type in Solar
  Cycles}}.
\bjtitle{\solphys}
\bvolume{292},
\bfpage{144}.
\doiurl{10.1007/s11207-017-1169-1}.
\adsurl{2017SoPh..292..144C}.
\end{barticle}
\endbibitem

\bibitem[\protect\citeauthoryear{Chinnery and North}{1975}]{Chinnery}
\begin{barticle}
\bauthor{\bsnm{Chinnery}, \binits{M.A.}},
\bauthor{\bsnm{North}, \binits{R.G.}}:
\byear{1975},
\batitle{The frequency of very large earthquakes}.
\bjtitle{Science}
\bvolume{190},
\bfpage{1197}.
\end{barticle}
\endbibitem

\bibitem[\protect\citeauthoryear{Clauset, Shalizi, and Newman}{2009}]{Clauset}
\begin{barticle}
\bauthor{\bsnm{Clauset}, \binits{A.}},
\bauthor{\bsnm{Shalizi}, \binits{C.R.}},
\bauthor{\bsnm{Newman}, \binits{M.E.}}:
\byear{2009},
\batitle{Power-law distributions in empirical data}.
\bjtitle{SIAM review}
\bvolume{51}(\bissue{4}),
\bfpage{661}.
\end{barticle}
\endbibitem

\bibitem[\protect\citeauthoryear{{Cliver} and
  {Dietrich}}{2013}]{2013JSWSC...3A..31C}
\begin{barticle}
\bauthor{\bsnm{{Cliver}}, \binits{E.W.}},
\bauthor{\bsnm{{Dietrich}}, \binits{W.F.}}:
\byear{2013},
\batitle{{The 1859 space weather event revisited: limits of extreme activity}}.
\bjtitle{Journal of Space Weather and Space Climate}
\bvolume{3}(\bissue{27}),
\bfpage{A31}.
\doiurl{10.1051/swsc/2013053}.
\adsurl{2013JSWSC...3A..31C}.
\end{barticle}
\endbibitem

\bibitem[\protect\citeauthoryear{Coles}{2001}]{Coles}
\begin{bbook}
\bauthor{\bsnm{Coles}, \binits{S.}}:
\byear{2001},
\bbtitle{An introduction to statistical modeling of extreme values},
\bsertitle{Springer Series in Statistics},
\bpublisher{Springer},
\blocation{Bristol UK}.
\end{bbook}
\endbibitem

\bibitem[\protect\citeauthoryear{de~Haan and Ferreira}{2007}]{deHaan}
\begin{bbook}
\bauthor{\bparticle{de} \bsnm{Haan}, \binits{L.}},
\bauthor{\bsnm{Ferreira}, \binits{A.}}:
\byear{2007},
\bbtitle{Extreme value theory: An introduction},
\bsertitle{Springer Series in Operations Research and Financial Engineering},
\bpublisher{Springer},
\blocation{New York}.
\bisbn{9780387344713}.
\end{bbook}
\endbibitem

\bibitem[\protect\citeauthoryear{Diaconis and Efron}{1983}]{Diaconis}
\begin{barticle}
\bauthor{\bsnm{Diaconis}, \binits{P.}},
\bauthor{\bsnm{Efron}, \binits{B.}}:
\byear{1983},
\batitle{Computer-intensive methods in statistics}.
\bjtitle{Scientific American}
\bvolume{248}(\bissue{5}),
\bfpage{116}.
\end{barticle}
\endbibitem

\bibitem[\protect\citeauthoryear{DiCiccio and Efron}{1996}]{DiCiccio}
\begin{botherref}
\oauthor{\bsnm{DiCiccio}, \binits{T.J.}},
\oauthor{\bsnm{Efron}, \binits{B.}}:
1996,
Bootstrap confidence intervals.
\textit{Statistical science},
189.
\end{botherref}
\endbibitem

\bibitem[\protect\citeauthoryear{{Gold} and
  {Hoyle}}{1960}]{1960MNRAS.120...89G}
\begin{barticle}
\bauthor{\bsnm{{Gold}}, \binits{T.}},
\bauthor{\bsnm{{Hoyle}}, \binits{F.}}:
\byear{1960},
\batitle{{On the origin of solar flares}}.
\bjtitle{\mnras}
\bvolume{120},
\bfpage{89}.
\doiurl{10.1093/mnras/120.2.89}.
\adsurl{1960MNRAS.120...89G}.
\end{barticle}
\endbibitem

\bibitem[\protect\citeauthoryear{Grabchak}{2016}]{Grabchak}
\begin{bchapter}
\bauthor{\bsnm{Grabchak}, \binits{M.}}:
\byear{2016},
\bctitle{Tempered stable distributions}.
In: \bbtitle{Tempered Stable Distributions},
\bpublisher{Springer},
\blocation{Charlotte, NC, USA}.
\end{bchapter}
\endbibitem

\bibitem[\protect\citeauthoryear{Jonas and McCarron}{2016}]{SWE:SWE20303}
\begin{barticle}
\bauthor{\bsnm{Jonas}, \binits{S.}},
\bauthor{\bsnm{McCarron}, \binits{E.D.}}:
\byear{2016},
\batitle{White house releases national space weather strategy and action plan}.
\bjtitle{Space Weather}
\bvolume{14}(\bissue{2}),
\bfpage{54}.
\bcomment{2015SW001357}.
\doiurl{10.1002/2015SW001357}.
\burl{http://dx.doi.org/10.1002/2015SW001357}.
\end{barticle}
\endbibitem

\bibitem[\protect\citeauthoryear{{Kane}, {McTiernan}, and
  {Hurley}}{2005}]{2005A&A...433.1133K}
\begin{barticle}
\bauthor{\bsnm{{Kane}}, \binits{S.R.}},
\bauthor{\bsnm{{McTiernan}}, \binits{J.M.}},
\bauthor{\bsnm{{Hurley}}, \binits{K.}}:
\byear{2005},
\batitle{{Multispacecraft observations of the hard X-ray emission from the
  giant solar flare on 2003 November 4}}.
\bjtitle{\aap}
\bvolume{433},
\bfpage{1133}.
\doiurl{10.1051/0004-6361:20041875}.
\adsurl{2005A\%26A...433.1133K}.
\end{barticle}
\endbibitem

\bibitem[\protect\citeauthoryear{Katsova \textit{et~al.}}{2018}]{Katsova2018}
\begin{barticle}
\bauthor{\bsnm{Katsova}, \binits{M.M.}},
\bauthor{\bsnm{Kitchatinov}, \binits{L.L.}},
\bauthor{\bsnm{Livshits}, \binits{M.A.}},
\bauthor{\bsnm{Moss}, \binits{D.L.}},
\bauthor{\bsnm{Sokoloff}, \binits{D.D.}},
\bauthor{\bsnm{Usoskin}, \binits{I.G.}}:
\byear{2018},
\batitle{Can superflares occur on the sun? a view from dynamo theory}.
\bjtitle{Astronomy Reports}
\bvolume{62}(\bissue{1}),
\bfpage{72}.
\doiurl{10.1134/S106377291801002X}.
\burl{https://doi.org/10.1134/S106377291801002X}.
\end{barticle}
\endbibitem

\bibitem[\protect\citeauthoryear{{Koons}}{2001}]{Koons}
\begin{barticle}
\bauthor{\bsnm{{Koons}}, \binits{H.C.}}:
\byear{2001},
\batitle{{Statistical analysis of extreme values in space science}}.
\bjtitle{\jgr}
\bvolume{106},
\bfpage{10915}.
\doiurl{10.1029/2000JA000234}.
\adsurl{2001JGR...10610915K}.
\end{barticle}
\endbibitem

\bibitem[\protect\citeauthoryear{Leadbetter, Lindgren, and
  Rootz{\'e}n}{2012}]{Leadbetter}
\begin{bbook}
\bauthor{\bsnm{Leadbetter}, \binits{M.R.}},
\bauthor{\bsnm{Lindgren}, \binits{G.}},
\bauthor{\bsnm{Rootz{\'e}n}, \binits{H.}}:
\byear{2012},
\bbtitle{Extremes and related properties of random sequences and processes},
\bpublisher{Springer},
\blocation{New York}.
\end{bbook}
\endbibitem

\bibitem[\protect\citeauthoryear{Meerschaert, Roy, and
  Shao}{2012}]{Meerschaert}
\begin{barticle}
\bauthor{\bsnm{Meerschaert}, \binits{M.M.}},
\bauthor{\bsnm{Roy}, \binits{P.}},
\bauthor{\bsnm{Shao}, \binits{Q.}}:
\byear{2012},
\batitle{Parameter estimation for exponentially tempered power law
  distributions}.
\bjtitle{Communications in Statistics-Theory and Methods}
\bvolume{41}(\bissue{10}),
\bfpage{1839}.
\doiurl{10.1080/03610926.2011.552828}.
\end{barticle}
\endbibitem

\bibitem[\protect\citeauthoryear{Riley}{2012}]{Riley}
\begin{botherref}
\oauthor{\bsnm{Riley}, \binits{P.}}:
2012,
On the probability of occurrence of extreme space weather events.
\textit{Space Weather}
\textbf{10}(2).
\end{botherref}
\endbibitem

\bibitem[\protect\citeauthoryear{Stumpf and Porter}{2012}]{Stumpf}
\begin{barticle}
\bauthor{\bsnm{Stumpf}, \binits{M.P.H.}},
\bauthor{\bsnm{Porter}, \binits{M.A.}}:
\byear{2012},
\batitle{Critical truths about power laws}.
\bjtitle{Science}
\bvolume{335}(\bissue{6069}),
\bfpage{665}.
\doiurl{10.1126/science.1216142}.
\burl{http://science.sciencemag.org/content/335/6069/665}.
\end{barticle}
\endbibitem

\bibitem[\protect\citeauthoryear{{Toriumi}
  \textit{et~al.}}{2017}]{2017ApJ...834...56T}
\begin{barticle}
\bauthor{\bsnm{{Toriumi}}, \binits{S.}},
\bauthor{\bsnm{{Schrijver}}, \binits{C.J.}},
\bauthor{\bsnm{{Harra}}, \binits{L.K.}},
\bauthor{\bsnm{{Hudson}}, \binits{H.}},
\bauthor{\bsnm{{Nagashima}}, \binits{K.}}:
\byear{2017},
\batitle{{Magnetic Properties of Solar Active Regions That Govern Large Solar
  Flares and Eruptions}}.
\bjtitle{\apj}
\bvolume{834},
\bfpage{56}.
\doiurl{10.3847/1538-4357/834/1/56}.
\adsurl{2017ApJ...834...56T}.
\end{barticle}
\endbibitem

\bibitem[\protect\citeauthoryear{{Wheatland}}{2010}]{Wheatland}
\begin{barticle}
\bauthor{\bsnm{{Wheatland}}, \binits{M.S.}}:
\byear{2010},
\batitle{{Evidence for Departure from a Power-Law Flare Size Distribution for a
  Small Solar Active Region}}.
\bjtitle{\apj}
\bvolume{710},
\bfpage{1324}.
\doiurl{10.1088/0004-637X/710/2/1324}.
\adsurl{2010ApJ...710.1324W}.
\end{barticle}
\endbibitem

\end{thebibliography}

\IfFileExists{\jobname.bbl}{} {\typeout{}
\typeout{****************************************************}
\typeout{****************************************************}
\typeout{** Please run "bibtex \jobname" to obtain} \typeout{**
the bibliography and then re-run LaTeX} \typeout{** twice to fix
the references !}
\typeout{****************************************************}
\typeout{****************************************************}
\typeout{}}

%
\end{article} 

\end{document}